# Electrospun nanofibers of polyCD/PMAA polymers and their potential application as drug delivery system

Michele F. Oliveira [a], Diego Suarez [a], Júlio Cézar Barbosa Rocha [b], Alvaro Vianna Novaes de Carvalho Teixeira [b], Maria E. Cortés [c], Frederico B. De Sousa [d],*, Rubén D. Sinisterra [a]

[a] Departamento de Química, Instituto de Ciências Exatas, Universidade Federal de Minas Gerais (UFMG), Belo Horizonte, 31270-901 MG, Brazil
[b] Departamento de Física, Centro de Ciências Exatas e Tecnológicas, Universidade Federal de Viçosa (UFV), Viçosa, 36570-000 MG, Brazil
[c] Departamento de Odontologia Restauradora, Faculdade de Odontologia, Universidade Federal de Minas Gerais (UFMG), Belo Horizonte, 31270-901 MG, Brazil
[d] Instituto de Física e Química, Universidade Federal de Itajubá (UNIFEI), Itajubá, 37500-903 MG, Brazil

*Keywords:*
Electrospinnin
Polymeric nanofibers
Poly-cyclodextrin
Drug delivery system
Propranolol

## ABSTRACT

Herein, we used an electrospinning process to develop highly efficacious and hydrophobic coaxial nanofibers based on poly-cyclodextrin (polyCD) associated with poly(methacrylic acid) (PMAA) that combines polymeric and supramolecular features for modulating the release of the hydrophilic drug, propranolol hydrochloride (PROP). For this purpose, polyCD was synthesized and characterized, and its biocompatibility was assessed using fibroblast cytotoxicity tests. Moreover, the interactions between the guest PROP molecule and both polyCD and βCD were found to be spontaneous. Subsequently, PROP was encapsulated in uniaxial and coaxial polyCD/PMAA nanofibers. A lower PROP burst effect (reduction of approximately 50%) and higher modulation were observed from the coaxial than from the uniaxial fibers. Thus, the coaxial nanofibers could potentially be a useful strategy for developing a controlled release system for hydrophilic molecules.

## 1. Introduction

Technologies associated with the development of drug delivery systems (DDS) have significantly increased in recent decades [1, 2]. DDS based on polymers have been widely used due to their considerable therapeutic efficacy and low side effects [3]. The fusion between polymer science and innovative processing techniques has led to new architectures with desired hierarchical structures and multiple functionalities for biomedical applications [4, 5]. In this sense, polymer fibers have attracted great interest, including for use as DDS, due to their typical properties, e.g., large surface area-to-volume ratio and possible surface modifications [6–8]. Moreover, drugs loaded in polymeric fibers can provide systemic and locoregional therapies compared with other DDS, such as nanoparticles, nanocapsules or micellar systems, which have intrinsic fluidity and are difficult to keep localized in a specific area of the body [9, 10].

Electrospinning is a simple and versatile technique that is capable of manufacturing continuous fibers with diameters ranging from micrometers down to several nanometers by applying strong electric fields, and this technique can be a useful alternative for pharmaceutical applications in which drugs incorporated in a polymeric solution or melt are used [11–13]. Fibers produced by electrospinning can combine different, natural and synthetic polymers, thereby exhibiting distinct and complementary functions [9, 14, 15]. Indeed, biocompatible polymers have been used by the pharmaceutical industry and have been approved by the FDA, such as polymethacrylates, which are widely applied as film-coating agents, as well as transdermal films, buccal patches and other devices [16]. This might be an interesting strategy for producing electrospun fibers for use as drug delivery systems.

Hence, we are comparing the release of the hydrophilic drug, propranolol hydrochloride (PROP), using two strategies: using uniaxial fibers and using coaxial fibers which combine poly(methacrylic acid) (PMMA) and poly-cyclodextrin (polyCD). The latter provides the host:guest properties, thereby providing many cavities for drug inclusion and polymeric features (high molecular weight) through chemically linked cyclodextrins (CDs, Fig. 1). Furthermore, CDs have been used to enhance pharmaceutical properties, leading to a modified solubility, stability, greater bioavailability and reduction in side effects; therefore, CDs are promising molecules for constructing advanced delivery systems [17–19]. Another important role is that the CDs presented in the polymer main chain can play in this system and this is the potential of CDs allowing it to be used as a crosslinking agent to improve the hydrophobicity of acrylic polymers, according to data reported in the literature [20].

PROP is a nonselective beta-blocker that is primarily used in the treatment of angina pectoris, cardiac arrhythmias, hypertension and many other cardiovascular disorders. PROP is well absorbed in the gastrointestinal tract, but it has a relatively low oral bioavailability (15–23%) because of extensive hepatic first-pass metabolism. In addition, PROP possesses physicochemical properties such as high solubility in water (50 mg mL$^{-1}$), a short half-life (3–5 h) and a low molecular weight (295.81 g mol$^{-1}$) that make it a suitable candidate for incorporation in DDS that use other routes of administration instead of oral administration [21].

* Corresponding author at: Instituto de Física e Química, Universidade Federal de Itajubá (UNIFEI), Itajubá, 37500-903 MG, Brazil.
E-mail addresses: fredbsousa@gmail.com, fredbsousa@unifei.edu.br (F.B. De Sousa).

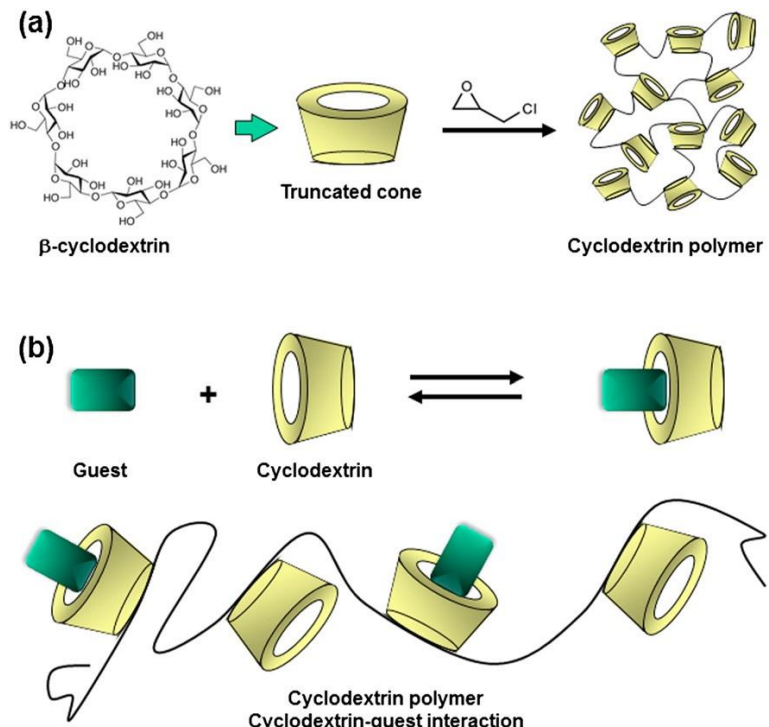

**Fig. 1.** (a) Chemical structure of βCD and a schematic representation of its truncated cone-shape and the reaction with epichlorohydrin to produce a CD-based polymer network; (b) comparison between cyclodextrin and polyCD host–guest interactions.

Controlled release of hydrophilic drugs loaded in polymeric fibers produced using conventional electrospinning techniques is still a challenge because their distribution on the fibers' surface can often result in rapid diffusion and lead to an increased burst release [22]. To obtain more efficacious release systems for highly soluble drugs, complex fiber arrangements, such as coaxial fibers and three-dimensional (3D) architectures, can be obtained using specific experimental conditions [9, 23, 24]. In coaxial electrospinning, two different polymer solutions are simultaneously pumped through a coaxial capillary and the drug is directly incorporated in the core, which remains protected by the shell. Thus, the drug release depends on both the core/shell polymers, which might promote a higher delivery modulation [25, 26].

Therefore, we report the preparation and characterization of uniaxial and coaxial PMAA/polyCD nanofibers for obtaining a more efficacious release system for the hydrophilic model drug, propranolol hydrochloride. First of all, polyCD was obtained and characterized, and then its supramolecular interaction with PROP was evaluated and compared with βCD, in order to identify the supramolecular complex structure and thermodynamic parameters. Moreover, we investigated the in vitro PROP release of uniaxial and coaxial nanofibers obtained by electrospinning combining polyCD and PMAA, after their fiber mats' characterization.

## 2. Materials and methods

### 2.1. Reagents and materials

Propranolol hydrochloride (PROP) was purchased from the Changzhou Yabang Pharmaceutical Co., Ltd.; β-cyclodextrin (βCD) was purchased from Xiamem Mchem Pharma Ltd.; poly(methacrylic acid) (PMAA), with a molecular weight of 100 kDa, was purchased from Polyscience, Inc.; and N,N-dimethylformamide ($C_3H_7NO$) and epichlorohydrin ($C_3H_5ClO$) were purchased from Sigma Aldrich. MTT (3-(4,5-dimethylthiazolyl-2)-2,5-diphenyltetrazolium bromide) was supplied by Invitrogen, and Dulbecco's Modified Eagle's Medium high-glucose (DMEM) was supplied by Sigma-Aldrich. Other reagents were of analytical grade and were used as received.

### 2.2. CD-based polymer and its characterization

#### 2.2.1. Polymer synthesis

PolyCD was synthesized according to methods previously described in the literature [27, 28] using βCD and EP. For this purpose, 10 g (8.8 mmol) of βCD was dissolved in 15 mL of a 15 wt.% sodium hydroxide aqueous solution, and the mixture was continuously stirred at 35 °C for 2 h. Then, 7 mL (88.0 mmol) of epichlorohydrin was added at a 1:10 βCD:EP molar ratio. The reaction was stopped after 4 h, which was before the gelation point, via the addition of acetone. Subsequently, the acetone was removed, and the pH of the aqueous solution was neutralized with a 6 mol L$^{-1}$ hydrochloric acid solution. The product was dialyzed for 7 days (molecular weight cut-off of 7000 kDa), and then the water was evaporated under vacuum at 60 °C to obtain the dry solid material.

#### 2.2.2. Nuclear magnetic resonance

NMR spectra were acquired using a Bruker DPX-400 AVANCE operating at 400 MHz at 27 °C with D2O (Cambridge isotopic 99.9%) as the solvent. The content of βCD in the polyCD was determined by 1H NMR.

#### 2.2.3. Light scattering

Static (SLS) and dynamic light scattering (DLS) measurements were performed using an apparatus from Brookhaven Co. and a He–Ne laser (Melles-Griot) with a wavelength of 632.8 nm. The temporal autocorrelation function of the scattered intensity was obtained at scattering angles ranging from 30 to 130°. DLS data were collected using a 1.0 wt.% polyCD solution. The increase in the refractive index of the polyCD ($dn/dC$) was directly determined using a differential refractometer (Brookhaven Co.) with polyCD solutions ranging in concentration from 1 to 10 mg mL$^{-1}$ and water as a reference. SLS data were collected using polyCD solutions with concentrations ranging from 1.48 to 4.76 mg mL$^{-1}$, and the same range of scattering angles was used for the SLS measurements.

### 2.2.4. Cell cytotoxicity

Cytotoxicity of polyCD was evaluated using an MTT assay, as described in the literature [29]. Immortalized human gingival fibroblasts (FMM1) were cultured in DMEM high-glucose medium supplemented with 10% fetal bovine serum and antibiotics (0.1 mg mL$^{-1}$ streptomycin and 100 U mL$^{-1}$ penicillin) and then incubated at 37 °C in a humidified atmosphere of 95% air and 5% $CO_2$. Upon reaching confluence, the cells were split, aliquoted ($9 \times 10^5$/cells per well) into 96-well plates and exposed for 48 h to polyCD solutions with a broad concentration range from 1.56 to $1.00 \times 10^4$ μg mL$^{-1}$. Subsequently, 60 μL of MTT was added to each well, and after 4 h, the formed salts were solubilized to formazan via the addition of SDS. Optical density measurements were performed at 570 nm using a Thermo Scientific Multiskan Spectrum MCC/340 spectrophotometer. Data are reported as the mean and standard deviation for six replicates for each concentration. Statistical analysis was performed using ANOVA, and $p < 0.05$ was considered to be statistically significant.

## 2.3. Supramolecular guest interaction with CD and polyCD

### 2.3.1. Nuclear magnetic resonance

PROP:βCD inclusion complexes were evaluated via 2D-ROESY measurements using the inversion–recovery sequence (90–t–180) with a mixing time of 600 ms. The water signal was used as the reference in all experiments.

### 2.3.2. Isothermal titration calorimetry

Isothermal titration calorimetry (ITC) was conducted using a TA Instruments NanoITC 2G at 298.15 K to access the thermodynamic parameters for the molecular interactions between the PROP:βCD and PROP:polyCD systems. Each titration consisted of 49 successive injections of 5 μL of a PROP aqueous solution (80 mmol L$^{-1}$) into the calorimetric cell that contained 1.0 mL of βCD (1 mmol L$^{-1}$) or polyCD aqueous solution (1 mmol L$^{-1}$ of βCD). Time intervals of 500 s were used to allow the signal to return to the baseline, and constant stirring at 250 rpm was kept constant during the experiment. Dilution processes were evaluated through the titration of βCD, polyCD and PROP in pure water (blank experiment) and were subtracted from the PROP:βCD and PROP:polyCD titration experiments. Data were analyzed using the software supplied with the instrument (NanoAnalyze software), and nonlinear regression (*independent fitting* model) was used to determine the binding constant ($K$), stoichiometry ($n$) and enthalpic contribution ($\Delta H$). Subsequently, the Gibbs free energy ($\Delta G$) and entropic contribution ($T\Delta S$) were calculated using thermodynamic equations described below:

$$\Delta G = -RT \ln K \quad (1)$$
$$\Delta G = \Delta H - T\Delta S. \quad (2)$$

## 2.4. Electrospinning process and fibers characterization

### 2.4.1. Electrospinning set up

To obtain uniaxial fibers, blend solutions were prepared using a total polymer concentration of 250 mg mL$^{-1}$ in DMF with overnight stirring, and the PMAA:polyCD ratios were 100:0, 80:20 and 60:40 wt.%, with 5 mg mL$^{-1}$ PROP. These blends were electrospun using a conventional electrospinning setup in which one solution passes through a single capillary assisted by a syringe pump (Harvard Apparatus). A solution flow rate of 3.0–2.5 mL h$^{-1}$, capillary tip-to-collector distance of 25 cm and voltage of ~15 kV were used during the electrospinning process. A special arrangement [30] was constructed to obtain coaxial fibers, in which individual polymer solutions were pumped by two coaxial capillaries supplied with the shell solution (PMAA solution) around the core solution (polyCD solution in addition to PROP). Shell and core solution flow rates were maintained constant at 2.0 and 1.0 mL h$^{-1}$, respectively, and the other parameters were the same as those used for electrospinning of the uniaxial fibers. The entire amount of PROP was considered to be incorporated into the fiber mats because a homogeneous polymer solution was obtained and the solution was completely electrospun. Subsequently, the fibers were annealed in an oven at 170 °C for 48 h to increase the hydrophobicity of the fibers through the formation of crosslinks between PMAA and polyCD.

### 2.4.2. Scanning and transmission electron microscopy

Surface morphologies of the fibers were investigated using scanning electron microscopy with a FEG-QUANTA 200 FEI at an accelerating voltage of 20 kV. Prior to obtaining the SEM images, all of the samples were coated with a 5 nm thick layer of gold using a sputter coater. Average fiber diameters were determined from at least 10 measurements in 3 different micrographs using the image analysis software ImageJ. The structures of the coaxial fibers were observed using transmission electron microscopy with a Tecnai G2-20–SuperTwin FEI operating at 200 kV. Samples for the TEM observations were prepared by directly depositing a thin layer of electrospun fibers on copper grids.

### 2.4.3. Fibers' wettability

The degree of wetting was performed to determine the hydrophilic/hydrophobic properties of the uni- and coaxial fibers via the sessile drop method using a video-based contact angle instrument in a KRUSS GmbH EasyDrop. Samples were cut and placed on the testing plate, and then distilled water (ten drops containing 10 μL each) was carefully dropped on the surfaces. Temporal images were generated from a computer analysis of the acquired images.

### 2.4.4. Attenuated total reflectance Fourier transform infrared spectroscopy

Spectra of the fiber surfaces were recorded using a Perkin Elmer Spectrum 100 IR spectrophotometer equipped with a universal ATR sampling accessory with a diamond top plate. Spectra were obtained with 128 scans per sample at a resolution of 4 cm$^{-1}$ between 4000 and 650 cm$^{-1}$. Spectra were processed using the software supplied with the instrument (Spectrum software).

### 2.4.5. In vitro drug release

PROP-loaded electrospun fibers (100 mg) were placed in 3 mL of phosphate-buffered saline (PBS, pH 7.2). The test was performed at 37 °C in an incubator-shaker at 50 rpm. Supernatant was completely removed at the selected intervals and replenished with an equal volume of fresh buffer solution. The concentrations of PROP were determined using a UV–vis spectrophotometer (Thermo Scientific Multiskan Spectrum MCC/340) at 290 nm. The employed PROP working range was 5.0–66.0 μg mL$^{-1}$, and a calibration curve was prepared for each set of measurements (correlation coefficient N 0.99). Each sample was assayed in triplicate, and the error bars show the standard deviation.

## 3. Results and discussion

### 3.1. Characterization of CD-based polymer

PolyCD was obtained via the polycondensation of βCD and epichlorohydrin, a bifunctional coupling agent, under strong alkaline conditions with a maximum yield of 41%. βCD content was determined by $^1$H NMR, considering that the glucopyranose ring spectrum shows a signal at δ 5.11 assigned to the anomeric proton H1 and that the two other signals are related to hydrogen atoms H2/H4 and H3/H5/H6 at δ 3.72 and 4.05, respectively, as shown in Fig. 2a [31]. After the polymerization process, an increase in the integration intensities of the aforementioned signals at δ 3.72 and 4.05 was observed due to the presence of five hydrogen atoms on one epichlorohydrin molecule, as shown in Fig. 2b. Integration peak ratio allows the degree of substitution to be determined, which was estimated to consist of 50 wt.% in βCD cavities.

In order to determine polyCD size and molecular weight, DLS and SLS experiments were carried out. DLS is a technique that allows the hydrodynamic radius (Rh) of a macromolecule to be calculated, which can be understood as the radius of a hypothetical hard sphere that has the same diffusivity of the particle being examined. Using the intensity autocorrelation function at 30° to 130° for the polyCD solution and the corresponding decay rate versus scattering wave vector plot shown in the supplementary data (Fig. SD 1), it was possible to obtain an $R_h$ value of 5.76 ± 0.03 nm using the Einstein–Stokes equation [32]:

$$R_h = \frac{k_B T}{6\pi\eta D} \quad (3)$$

where $k_B$ is the Boltzmann constant, $T$ is the absolute temperature, $\eta$ is the viscosity, and $D$ is the effective diffusion coefficient.

Multi-angle SLS is a convenient method for obtaining three important molecular parameters during a unique experiment: the weight-average molar mass, the radius of gyration and the second virial coefficient of macromolecules. These parameters can be determined through measurements of the intensity of light scattered under different concentrations and at various angles according to the Zimm equation, which is expressed as [33, 34]:

$$\frac{KC}{\Delta R_\theta} = \frac{1}{\bar{M}_w \left[1 + \frac{q^2}{3}\langle R_g^2 \rangle\right]} + 2A_2 C \quad (4)$$

where $K$ denotes the optical constant, $C$ is the concentration of polymer, $\Delta R_\theta$ is the Rayleigh ratio, $\bar{M}_w$ is the weight-average molar mass, $\langle R_g^2 \rangle$ is the mean square radius of gyration, $A_2$ is the second virial coefficient, and $q$ is the modulus of the scattering vector. Table 1 presents the $dn/dC$ value for polyCD and typical Zimm plot results. Supplementary data (Figs. SD 2 and SD 3) shows how the refractive indices of polyCD solutions vary for a given increase in concentration and the Zimm plot.

Zimm plot of polyCD shows that it possesses a high weight-average molar mass $\bar{M}_w$, which is a desirable physico-chemical characteristic for obtaining uniform fibers, as described elsewhere [35]. Previous studies have reported that the weight-average molar mass of polyCDs depends on the experimental conditions, such as the reaction time, the EP/βCD molar ratio, temperature and NaOH concentration [28]. In the present work, all of these parameters were controlled to develop a reproducible synthesis for polyCD.

The $R_g$ denotes the root-mean-square distance of an end from the center of gravity, which is an average measure of the size of the macromolecule. By combining the $R_g$ and $R_h$ values obtained using SLS and DSL techniques, it is possible to calculate the ratio $R_g/R_h$, also called the $\rho$ parameter, which indicates the morphology of the scatterers. The obtained $\rho$ value was 3.5 ± 0,5, which is higher than the expected value value for random coils ($\rho = 1.505$) but expected from very elongated structures, such as rigid rods. Positive $A_2$ values indicate favorable interactions between the polymer and solvent (water). An increase in this term was observed in comparison with the small negatives values reported for β-cyclodextrin [36] and this result can be attributed to a greater possibility of forming hydrogen bonds between the polymer and water. Similar systems formed by polyrotaxanes, which consisted of α-cyclodextrin and poly(ethylene glycol) (PEG), showed $A_2$ values with the same order of magnitude in different types of solvent systems [37].

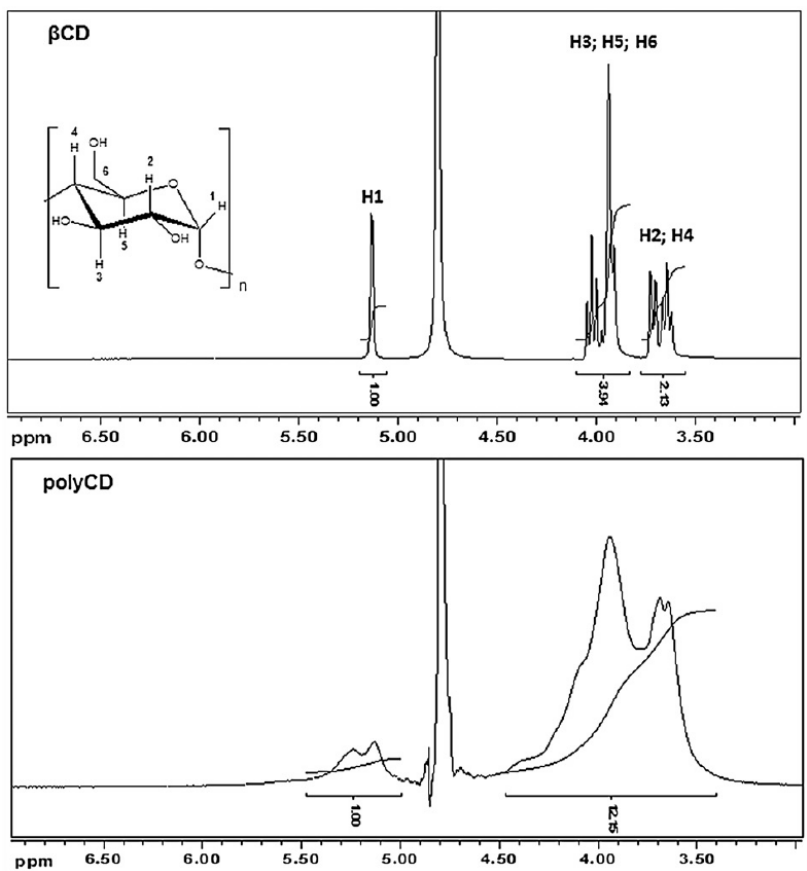

**Fig. 2.** $^1$H NMR spectra of (a) βCD and (b) polyCD at 400 MHz in D2O at 27 °C.

**Table 1**
$dn/dC$ value for polyCD and typical Zimm plot results.

| Polymer | $dn/dC$ (mL g$^{-1}$) | $\bar{M}_w$ (g mol$^{-1}$) | $R_g$ (nm) | $A_2$ (cm$^3$ mol g$^{-2}$) |
|---|---|---|---|---|
| PolyCD | $1.2394 \times 10^{-4}$ | $(6.0 \pm 0.1) \times 10^4$ | $(20 \pm 3)$ | $(1.5 \pm 0.3) \times 10^{-4}$ |

However, for DDS application, this polyCD should present low cytotoxicity. In the literature, diverse degrees of cytotoxicity related to different types of polyCD can be found [38–41]. To investigate the applicability of polyCD as a polymer matrix for drug delivery, the cytotoxicity was evaluated in vitro using fibroblasts. Cytotoxicity of polyCD was tested over a broad concentration range from 1.56 to $1.00 \times 10^4$ µg mL$^{-1}$ on the human fibroblast FMM1 cell line until 48 h, and results are presented in Fig. 3a and b. As observed, this polymer could be considered bio-compatible based on the low cytotoxicity observed. Thus, these results suggest that this polyCD polymer has considerable potential as a drug carrier. Moreover, this polyCD has greater biocompatibility compared to other polyCDs described in the literature and compared to the lower cytotoxicity of polyCD at 25 µg mL$^{-1}$ in comparison with the almost 100% cytotoxicity of a similar polyCD system at the same concentration [42].

### 3.2. Supramolecular guest interaction with CD and polyCD

To confirm the existence of intermolecular interactions between βCD and PROP and to determine the molecular orientation of the drug in the cavity of the CD, two-dimensional 2D-ROESY experiments were performed because this technique is one of the most effective techniques for studying cyclodextrin inclusion complexes. Fig. 4 shows the 2D-ROESY partial contour map and its expansion in D2O for the PROP:βCD system prepared via the freeze-drying method at 1:1 molar ratio, as described in previous works [17, 18]. Cross-peak correlations among the CD internal (H3 and H5) and external (H2 and H4) hydrogens and PROP aromatic hydrogens can be observed, indicating a short spatial distance between both molecules. This result confirms that the inclusion process of the aromatic region of the PROP molecule is preferentially inserted into the CD cavity as reported for other supramolecular systems [36].

Since the supramolecular structure was determined, ITC experiments were conducted to not only assess the thermodynamic parameters for the molecular interactions between PROP and βCD, but also to assess these parameters for the interactions between PROP and the polyCD polymeric system. These results are presented in Table 2 (see Fig. SD 4 for the titration curves).

Based on these titrations curves, it was possible to confirm that not only the host:guest interaction between the PROP:βCD but also that between the PROP:polyCD were spontaneous processes with favorable enthalpy and entropy contributions. Enthalpic contribution was associated with the release of water molecules from the βCD cavity to the bulk and with the intermolecular interactions between the host and guest molecules. Entropic contribution could be associated with the new conformation that was adopted due to the supramolecular interactions.

Surprisingly, a higher binding constant (K) was observed for the PROP:polyCD system than for the PROP:βCD supramolecular complex. This difference in the K constant could be understood based on the higher probability of βCD cavities on the polymer structure interacting with the PROP in comparison with the free βCD. In addition, this process may be due to the higher hydrogen bonding and van der Waals interactions. ITC results could also provide information about the molecular stoichiometries of both systems, in which more than one guest molecule interacts with a single βCD [43]. Similar superstructures have been observed for CD systems, and in these cases, the guest molecule acted as a glue between the inclusion complexes [36]. Other studies have described interactions between CD-based polymers and guest molecules and their use as a promising drug carriers using ITC [44, 45] These results strongly suggest that polyCD can be used as a polymer for DDS for hydrophilic molecules such as PROP.

### 3.3. Electrospinning process and fiber characterization

PMAA and PMAA:polyCD nanofibers were successfully obtained as uniaxial and coaxial nanofibers in the presence of PROP, and the SEM and TEM images obtained for these nanofibers before the annealing treatment are shown in Fig. 5. From the SEM images shown in Fig. 5a–h, we can observe that all nanofibers are randomly aligned (based on the electrospinning setup), bead free and have a relatively narrow distribution, as shown in Table 3. Moreover, note that the addition of PROP to the polymer mats did not affect the morphology or diameter of the nanofibers. A similar result was obtained for the addition of polyCD to the uniaxial nanofibers. Coaxial nanofibers (polyCD core and PMAA shell) present similar diameters in the presence and absence of PROP. However, the coaxial nanofibers are larger in diameter compared to the uniaxial mats, which is a consequence of the electrospinning setup. This can be confirmed by the TEM images, Fig. 5i–j, in which the core diameter is approximately 260 nm, corresponding to the uniaxial diameter. We could also identify the core-shell structure in the TEM images, which confirmed that the electrospinning setup was capable of producing the proposed mat structures. Morphologies of the fibers after the annealing treatment were evaluated using SEM, which did not reveal morphological changes (data not shown).

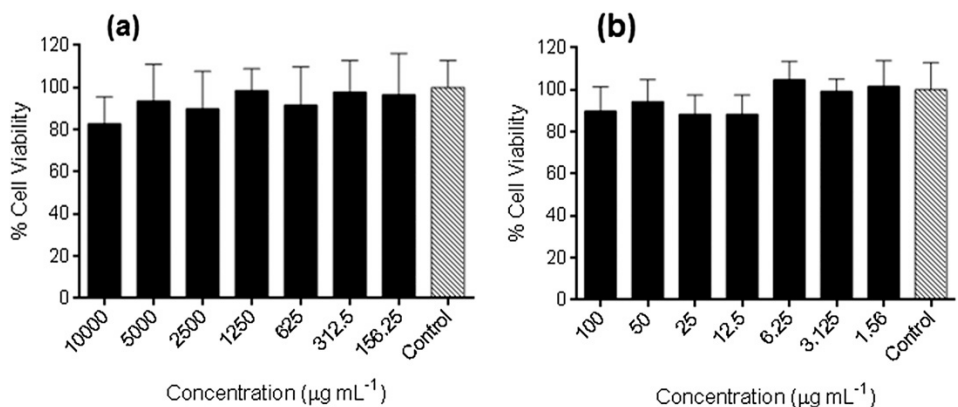

**Fig. 3.** Effect of polyCD on the viability of fibroblasts cells: (a) concentration range from 1.56 to 100 µg mL$^{-1}$ and (b) concentration range from 156.25 to $1.00 \times 10^4$ µg mL$^{-1}$.

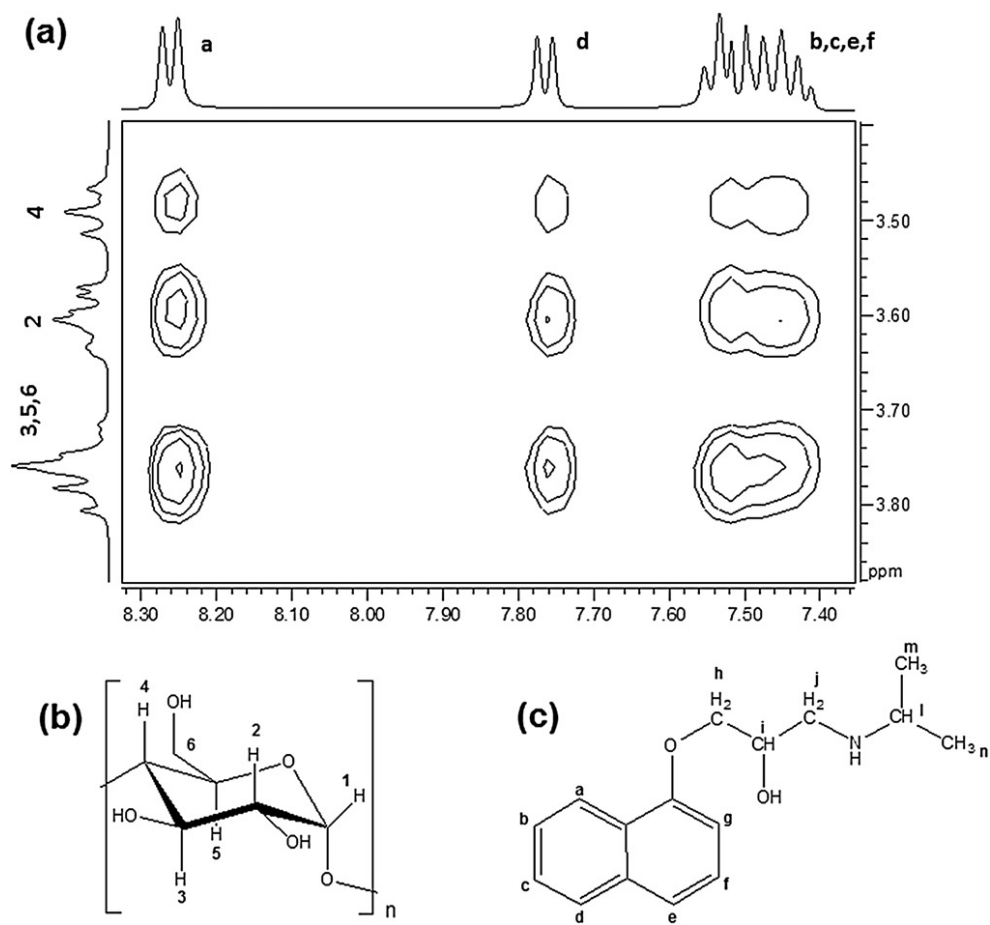

**Fig. 4.** (a) 2D-ROESY partial contour map at 400 MHz in D$_2$O for the PROP:βCD system with a 1:1 molar ratio, (b) cyclodextrin structure and (c) propranolol structure.

The annealing process at 170 °C was conducted to increase the hydrophobicity of the fibers through the formation of crosslinking between PMAA and polyCD, because thermal degradation of PROP was observed at 250 °C in the TG/DTG curves (see Fig. SD 5), and the 170 °C temperature could be used for all fibers. This treatment has already been described for other acrylic polymers using CDs as the crosslinking agent. The reaction mechanism involves the formation of a cyclic anhydride via the dehydration of carboxylic acid groups from the acrylic polymer. Subsequently, the anhydride reacts with the hydroxyl groups of CD, resulting in ester bonds that are responsible for the increase in the hydrophobicity of the polymer [20]. Fig. 6 shows the crosslinking process between PMAA and polyCD.

The annealing process was assessed using ATR-FTIR, and Fig. 7 shows the spectrum of PMAA fibers, which presents characteristic bands at $v_{max}$/cm$^{-1}$ 3400, which is a broad band corresponding to –OH stretching, and at 1697, 1393, and 960–930, in which the first band is related to free C=O stretching and the other ones are related to the acid dimer. In addition, after the annealing process, two new bands were observed at $v_{max}$/cm$^{-1}$ 1803 and 1021 as a result of the crosslink process between the carbonyl groups of the PMMA and the polyCD polymer, as described elsewhere [46].

In the uniaxial and coaxial fibers, polyCD bands are present at $v_{max}$/cm$^{-1}$ 3360 (OH stretching), 2921 (C–H stretching) and 1025 (C–O–C stretching), [47] and these bands were overlapped with those of the PMAA. These fibers also exhibited a band at $v_{max}$/cm$^{-1}$ 1803 and an increase in intensity of $v_{max}$/cm$^{-1}$ 1021, indicating the occurrence of a crosslink. However, in this case, this process could occur between the hydroxyl groups of polyCD and the carbonyl groups of PMAA [20].

Moreover, wettability of the PMAA, uniaxial PMAA:polyCD (80:20), and PMAA:polyCD (60:40) fibers in water after the annealing process were monitored as a function of time. It can be observed that the uniaxial fibers without PROP did not dissolve and that the water was immediately adsorbed, showing high affinity for these surfaces. These results demonstrated the effectiveness of the annealing treatment and might improve their use as a drug carrier.

Uniaxial fibers exhibit a surface property that depends on the polyCD concentration in the presence of PROP, as shown Fig. SD 6. A higher time to adsorb the water drop (approximately 300 s) by PMAA:polyCD (60:40) nanofibers was observed in comparison with the other fibers, as shown in Table 3. This result suggests that PROP interacts with polyCD to produce a more hydrophobic compound, most likely based on the supramolecular interactions between the cavities of CD (and the hydroxyl groups of CD) and the guest molecule. Coaxial fibers presented a result similar to that of the uniaxial fibers without PROP, indicating that PROP and polyCD are present in the core of the fiber mat, corroborating the TEM image and suggesting a more hydrophobic matrix for delivery of the hydrophilic drug PROP.

**Table 2**
Thermodynamic parameters for supramolecular interactions between PROP:βCD and PROP:polyCD.

| Systems | ΔG (kJ mol$^{-1}$) | ΔH (kJ mol$^{-1}$) | TΔS (kJ mol$^{-1}$) | K | n |
|---|---|---|---|---|---|
| PolyCD | −14.9 | −3.4 ± 0.6 | 11.5 | 408.3 ± 47.6 | 2.8 ± 0.4 |
| PROP: polyCD | −17.3 | −2.1 ± 0.5 | 15.2 | 1075.7 ± 318.6 | 3.5 ± 0.5 |

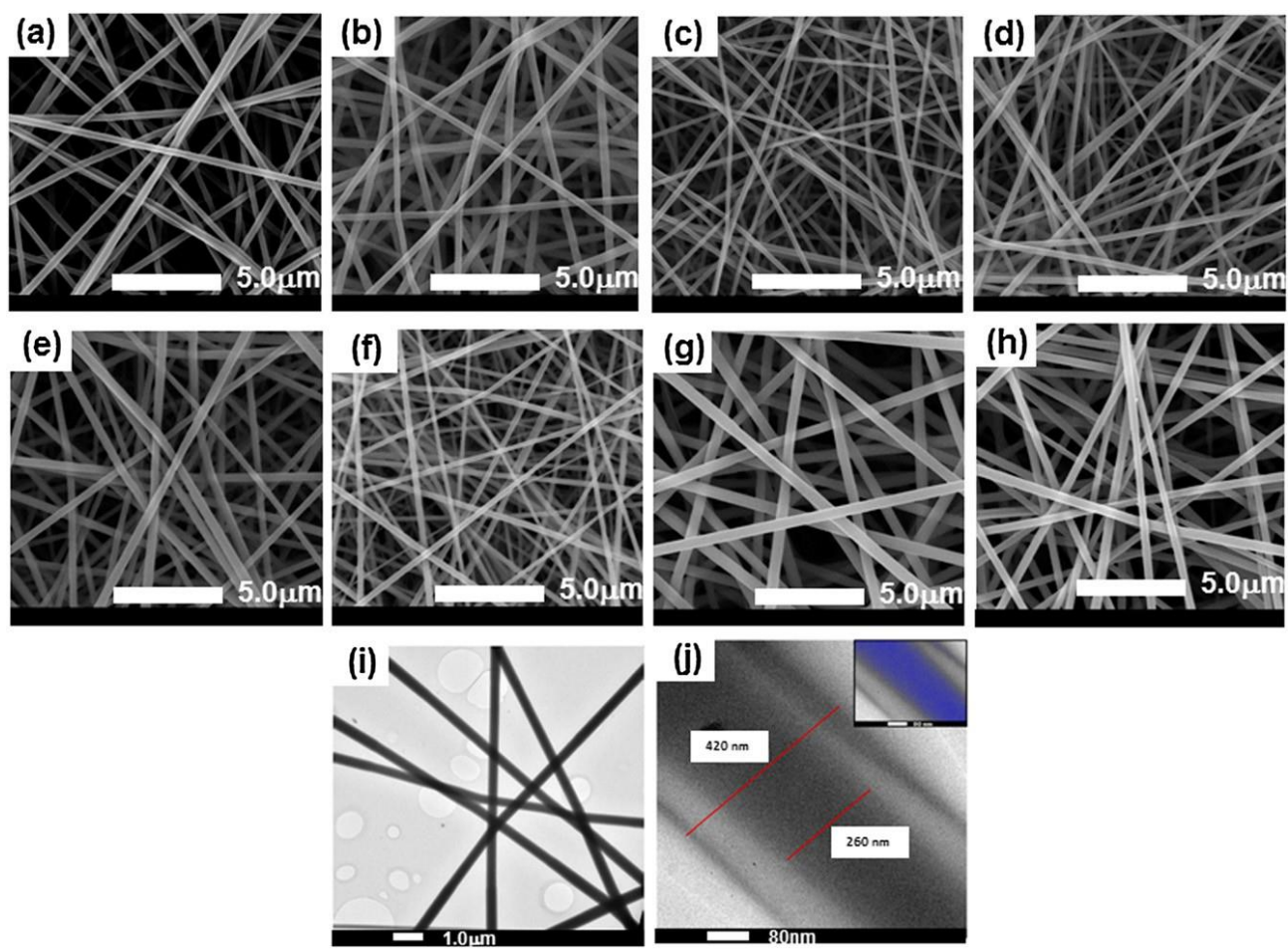

**Fig. 5.** SEM (magnification 10.000×) and TEM images of uni- and coaxial PMAA:polyCD fibers: (a) SEM uniaxial–PMAA; (b) SEM uniaxial–PMAA + PROP; (c) SEM uniaxial PMAA:polyCD (80:20); (d) SEM uniaxial PMAA:polyCD (80:20) + PROP; (e) SEM uniaxial PMAA:polyCD (60:40); (f) SEM uniaxial PMAA:polyCD (60:40) + PROP; (g) SEM coaxial–shell (PMAA) and core (polyCD); (h) SEM coaxial–shell (PMAA) and core (polyCD + PROP); (i) TEM coaxial–shell (PMAA) and core (polyCD + PROP) (magnification 10.000×); and (j) TEM coaxial–shell (PMAA) and core (polyCD + PROP) (magnification 230.000×).

Incorporation of polyCD into PMAA to prepare polymeric nanofibers was proposed to control the release of PROP to evaluate how different structures (uni- and coaxial fibers) can affect this process. In similar systems, it is already known that drug release depends on the drug solubility, crosslink network and supramolecular interactions with the CD cavities [45], which were also demonstrated above through ITC experiments with our system. Recently, Thatiparti et al., prepared

**Table 3**
Diameters of uniaxial and coaxial PMAA and PMAA:polyCD nanofibers obtained from scanning and transmission electron microscopy images and water adsorption time obtained by con- tact angle measurements.

| Fibers | PMAA:polyCD ratio/wt.% | Diameter/nm | Water adsorption time/s |
|---|---|---|---|
| Uniaxial (PMAA) | 100:0 | (290 ± 35) | a |
| Uniaxial (PMAA + PROP) | 100:0 | (310 ± 38) | 60 |
| Uniaxial PMAA:polyCD blends | 80:20 | (254 ± 45) | a |
|  | 60:40 | (305 ± 45) | a |
| Uniaxial PMAA:polyCD blends + PROP | 80:20 | (252 ± 37) | 120 |
|  | 60:40 | (250 ± 34) | 300 |
| Coaxial PMAA:polyCD | 100:0 (shell) 0:100 (core) | (418 ± 54) | a |
| Coaxial PMAA:polyCD + PROP | 100:0 (shell) 0:100 (core) | (404 ± 72) | a |

[a] Water drop immediately adsorbed.

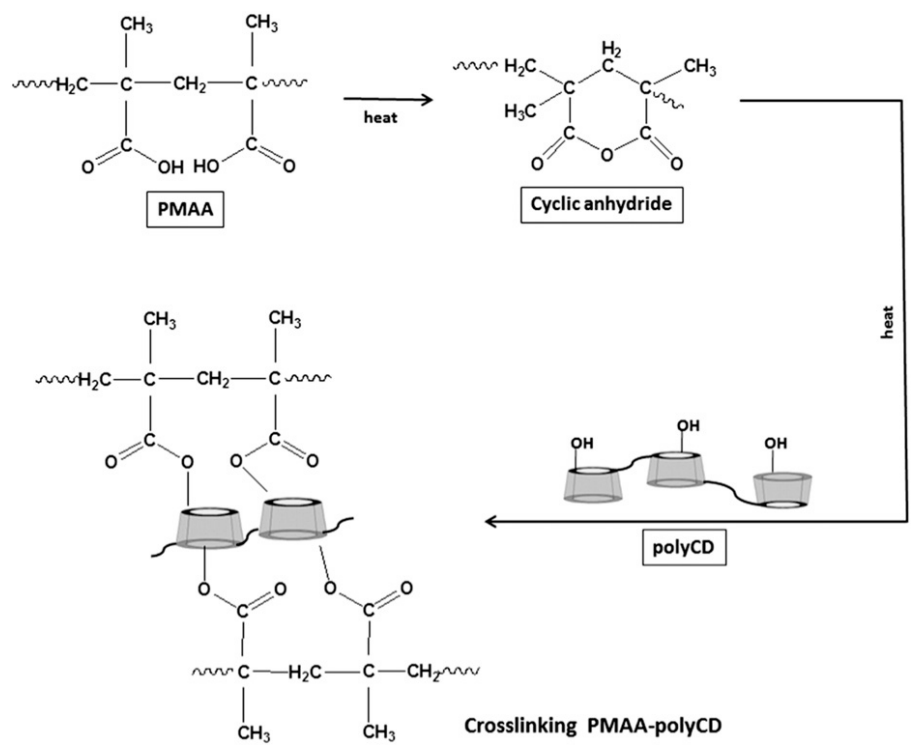

**Fig. 6.** Crosslinking process between PMAA and polyCD.

CD-based polymers in which diisocyanates were used as a coupling agent toevaluate them as a platform for delivering antibiotics. It was observed that the release of drugs from the CD-based gels was slower than that from dextran gels (used for comparison) and that the release could be sustained for more than 200 h. In addition, these systems showed greater bactericidal activity against Staphylococcus aureus, reflecting their potential as a delivery system. Another study conducted by Garcia-Fernandez et al. [48] incorporated ethoxzolamide, a drug applied in the treatment of glaucoma, into CD-based polymers for soft contact lenses based on poly(2-hydroxyethyl methacrylate). Authors observed that the CD-based polymers facilitated the loading of high doses of drug into contact lenses and led to the retention of the drug, providing a sustained release for several weeks.

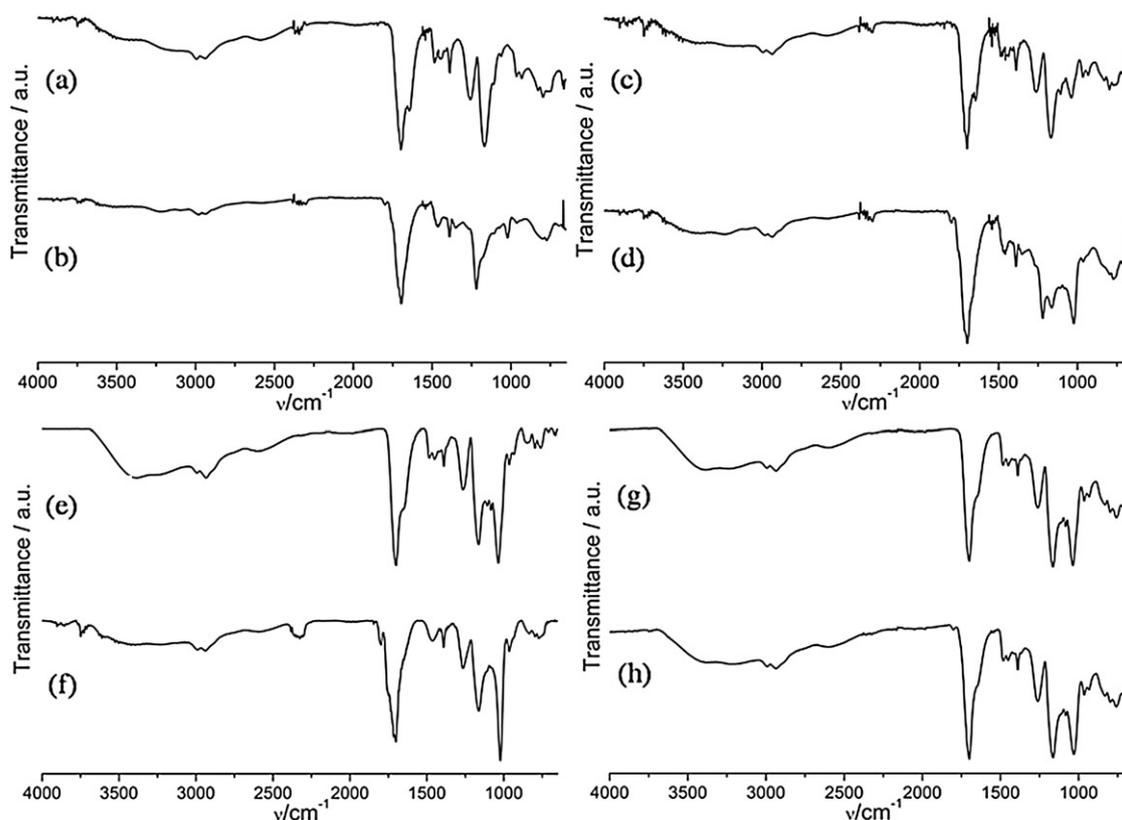

**Fig. 7.** ATR-FTIR spectrum for PMAA:polyCD fibers in the range of 4000–650 cm$^{-1}$ before the annealing process: (a) uniaxial PMAA + PROP; (c) uniaxial PMAA:polyCD (80:20) + PROP;(e) uniaxial PMAA:polyCD (60:40) + PROP; (g) coaxial–shell (PMAA) and core (polyCD + PROP) and after annealing process: (b) uniaxial PMAA + PROP; (d) uniaxial PMAA:polyCD (80:20) + PROP; (f) uniaxial PMAA:polyCD (60:40) + PROP; and (h) coaxial–shell (PMAA) and core (polyCD + PROP).

Herein, the release profiles of PROP from fibers were evaluated until 168 h and are illustrated in Fig. 8a. Data for PROP-loaded PMAA fibers are not shown because these fibers were dissolved after 15 min upon contact with buffer solution; thus, approximately 100% of PROP was released, and the fibers did not exhibit control over the drug release. Higher PROP releases of 30 and 35% from the PMAA:polyCD (80:20) and (60:40) uniaxial fiber matrices, respectively, in the first 8 h were observed. These results are in contrast to the expected result of the more hydrophobic PMAA:polyCD (60:40) matrix presenting a lower propranolol release, but only a slight difference of 5% in the burst effect was observed between both matrices. This result could be due not only to the higher hydrophobicity of the nanofibers after the crosslinking process but also to the larger PMAA polymer causing greater steric hindrance as a result of the degree of ester bond formation, which causes greater difficulties for the propranolol complexation capacity in the PMAA:polyCD (60:40) uniaxial fibers [49]. Interestingly, a lower burst effect of 15% was observed for the coaxial fibers compared to the uniaxial fibers. This result could be explained by the higher probability of PROP chemical interactions with the fiber.

A higher PROP release of 40% was observed at 168 h from the uniaxial fibers compared to the release of 23% from the coaxial fibers. These results appear to be attributed to the presence of polyCD that assists in the crosslink process with PMAA, the higher hydrophobicity of polyCD than the PMAA fibers, and the interaction with PROP to form supramolecular systems, thereby delaying the release. Additionally, one could suggest that the highest percentage of polyCD present in the core of the coaxial fibers compared to the amount dispersed in uniaxial fibers can lead to the drug released from one cavity becoming available to form supramolecular interactions with empty CDs and delaying the release during their diffusion along the polymer matrices and decreasing the burst release effect of highly water soluble molecules.

Thus, the higher PROP release from uniaxial fibers might be due to the drug dispersion throughout the fibers, including on their surfaces. In fact, previous studies have proposed that using coaxial fibers is an interesting strategy for controlling drug release because the drug is incorporated into the polymers as the core and is not in direct contact with the medium. For instance Sohrabi et al. [50], designed a drug delivery system based on coaxial nanofibers of poly(methyl methacrylate)(PMMA)- nylon6 that contained ampicillin as a model drug. Authors observed that these systems were capable of releasing the drug in a sustained manner [51]. They also reported that a clear difference exists in the release profiles of hydrochloride metoclopramide when uniaxial fibers prepared with poly(ε-caprolactone) (PCL), poly(lactic acid) (PLA), poly(lactic-co-glycolic) (PLGA) and polyvinyl alcohol (PVA) were used and when their coaxial fibers were prepared using polyvinyl alcohol (PVA) as a core. Thus, we could suggest that the coaxial fibers could modulate the PROP release profile more efficiently, showing potential as a useful strategy for the release of hydrophilic drugs.

Finally, SEM images were obtained from the nanofibers used in the drug delivery system to evaluate the morphologies of the fibers after 168 h. From these images shown in Fig. 8b–d, a collapse of the uniaxial fibers and the formation of a rough film can be observed. In contrast, the structure of the coaxial fibers was retained during this period. These SEM results support the hypothesis that the structures of the nanofibers are also responsible for greater modulation of PROP release and that they can be employed as a device for drug delivery.

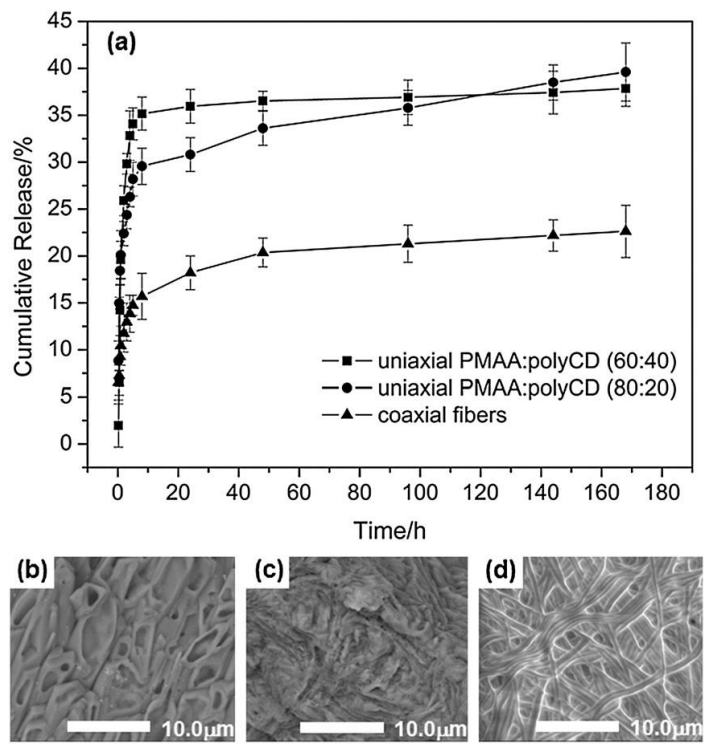

**Fig. 8.** (a) Release profiles of PROP from PMAA:polyCD fibers evaluated over 168 h and SEM images of these fibers after the release: (b) uniaxial fiber PMAA:polyCD (80:20) + PROP;(c) uniaxial fiber PMAA:polyCD (60:40) + PROP; and (d) coaxial fiber shell (PMAA) and core (polyCD + PROP).

## 4. Conclusions

In summary, we prepared nanofibers consisting of a CD-based biocompatible polymer (polyCD) and associated with poly(methacrylic acid) via electrospinning as a strategy for developing a drug delivery system for more hydrophilic drugs, and the hydrophilic drug PROP was used as the model drug. The synthesized polyCD with a high weight-average molar mass contained βCD cavities that were able to spontaneously encapsulate the drug through host–guest interactions. This system was successfully electrospun into uni- and coaxial randomly oriented nanofibers, and the polymer matrix exhibited biocompatibility. The annealing process between the polyCD and poly(methacrylic acid) was investigated and favored the formation of more hydrophobic fibers that could be used as an interesting drug delivery carrier. The burst effect release of the hydrophilic PROP was drastically modulated by the coaxial fibers compared with the uniaxial fibers. Thus, this type of coaxial nano-fiber based on polyCD and poly(methacrylic acid) could be a useful strategy for delivering hydrophilic drugs such as propranolol.

## Acknowledgments

The authors are grateful to the Coordenação de Aperfeiçoamento de Pessoal de Nível Superior (CAPES, Ph.D fellowship), Conselho Nacional de Desenvolvimento Científico e Tecnológico (CNPq, project 477529/2012-7), Fundação de Amparo à Pesquisa do Estado de Minas Gerais (FAPEMIG-RED-00010-14), Instituto Nacional de Ciência e Tecnologia Nanobiofar (INCT-MICT/CNPq-FAPEMIG project 573924/2008-2), and the Nanofar CNPq (project 564796/2010-7) network for financial support. The authors are very thankful to PhD candidates Savio Morato and João Paulo Trigueiro from the Dentistry and Chemistry Departments at UFMG for their help in cellular and TEM experiments respectively. This work is a collaboration research project of members of the Rede Mineira de Química (RQ-MG) supported by FAPEMIG.

## Appendix A. Supplementary Data

Supplementary data to this article can be found online at http://dx.doi.org/10.1016/j.msec.2015.04.042.

# Appendix A. Supplementary data

# Electrospun nanofibers of PolyCD/PMAA polymers and their potential application as drug delivery system

*Michele F. Oliveira[a], Diego Suarez[a], Júlio Cézar Barbosa Rocha[b], Alvaro Vianna Novaes de Carvalho Teixeira[b], Maria E. Cortés[c], Frederico B. De Sousa*[d] and Rubén D. Sinisterra[a]*

[a]*Departamento de Química, Instituto de Ciências Exatas, Universidade Federal de Minas Gerais (UFMG), Belo Horizonte, 31270-901, MG, Brazil.*

[b]*Departamento de Física, Centro de Ciências Exatas e Tecnológicas, Universidade Federal de Viçosa (UFV), Viçosa, 36570-000, MG, Brazil.*

[c]*Departamento de Odontologia Restauradora, Faculdade de Odontologia, Universidade Federal de Minas Gerais (UFMG), Belo Horizonte, 31270-901, MG, Brazil.*

[d]*Instituto de Física e Química, Universidade Federal de Itajubá (UNIFEI), Itajubá, 37500-903, MG, Brazil.*

Corresponding author: Frederico B. De Sousa
fredbsousa@gmail.com, fredbsousa@unifei.edu.br
Instituto de Física e Química, Universidade Federal de Itajubá (UNIFEI), Itajubá, 37500-903, MG, Brazil.

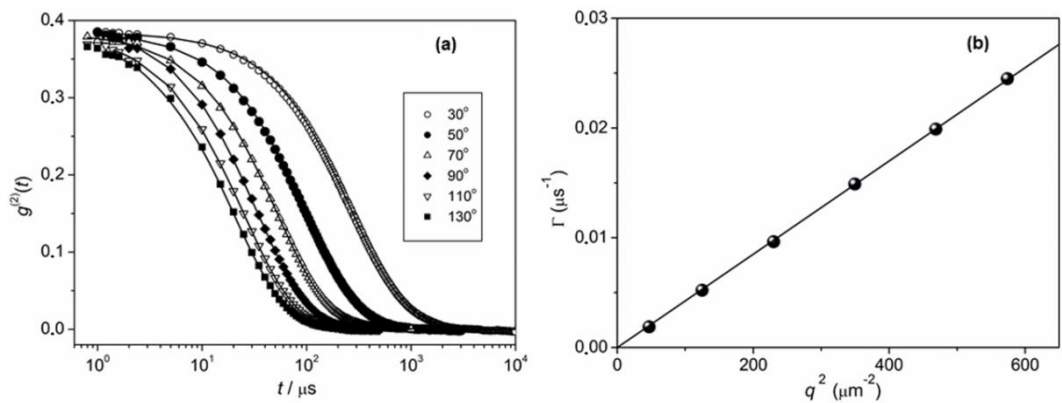

Fig. SD 1 (a) Intensity autocorrelation function at 30 to 130° for polyCD solution and (b) corresponding decay rate *versus* scattering wave vector plot for polyCD.

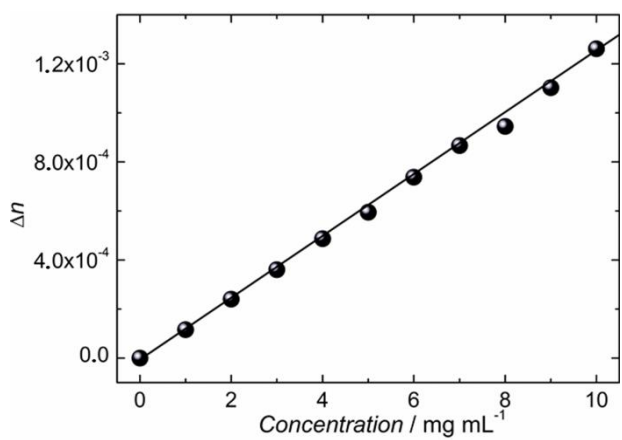

Fig. SD 2 Refractive index increment of polyCD.

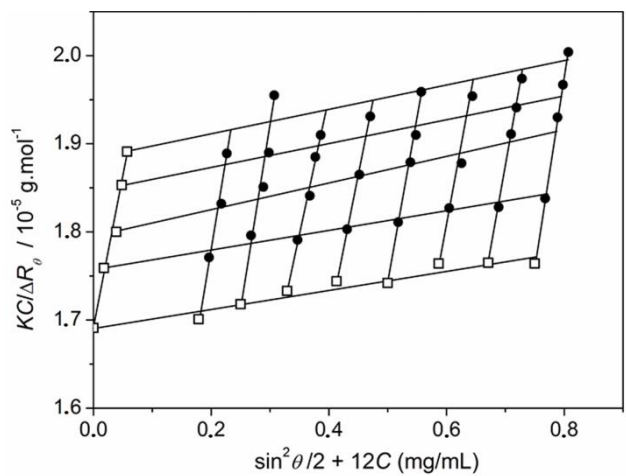

Fig. SD 3 Zimm plot for polyCD at 30 to 130°.

The Zimm equation:

$$\frac{KC}{\Delta R_\theta} = \frac{1}{\overline{M}_w\left[1+\frac{q^2}{3}\langle R_g^2\rangle\right]} + 2A_2C \tag{A.1}$$

The term $K$ is expressed by:

$$K = \frac{2\pi^2}{\lambda_0^4 N_A}\left(n_0\frac{dn}{dC}\right)^2 \tag{A.2}$$

which $\lambda_0$ is the light wavelength in the vacuum, $N_A$ is the Avogadro's number, $n_0$ is the refractive index of the solvent and $dn/dC$ is the refractive index increment of the solute with respect to the solvent.

Values of $\overline{M}_w$, $R_g$ $(= \langle R_g^2\rangle^{1/2})$ and $A_2$ were obtained through the construction the Zimm plot: $KC/R_\theta$ vs. $\sin^2(\theta/2) + kC$ by extrapolation to $C = 0$ and $\theta = 0$, where $k$ is an arbitrary constant to adjust the size of the plot. The term $1/\overline{M}_w$ was determined by intersecting with the ordinate axis, $R_g$ and $A_2$ by the slopes plot.

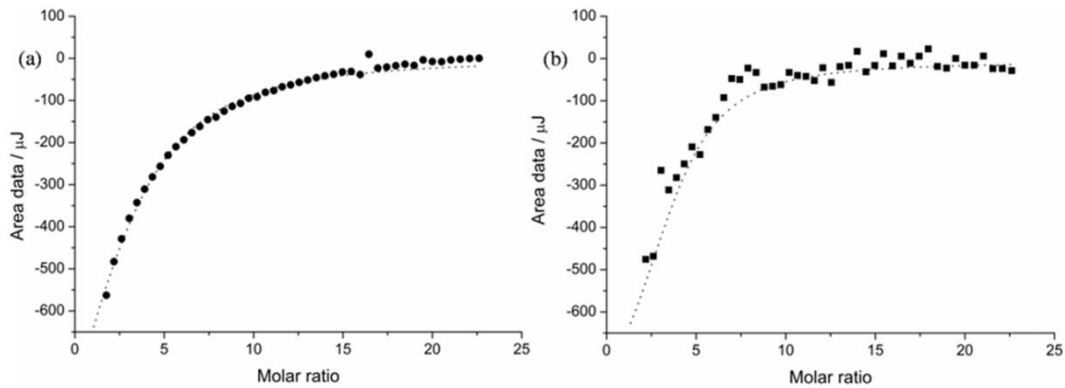

Fig. SD 4 Titration curves at 298.15 K for (●) PROP (80.0 mmol L$^{-1}$) in βCD (1.0 mmol L$^{-1}$) and (■)PROP (80.0 mmol L$^{-1}$) in polyCD (1 mmol L$^{-1}$ of βCD) and (·····) *independent fitting* model.

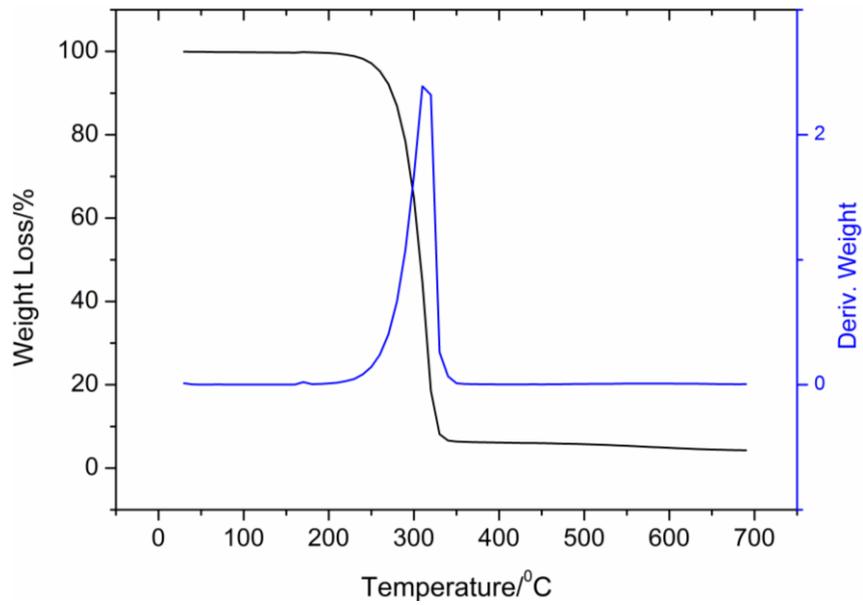

Fig. SD 5 TG/DTG curves for propranolol hydrochloride from room temperature to 700 °C at a rate of 10 °C min$^{-1}$ under a nitrogen gas flow.

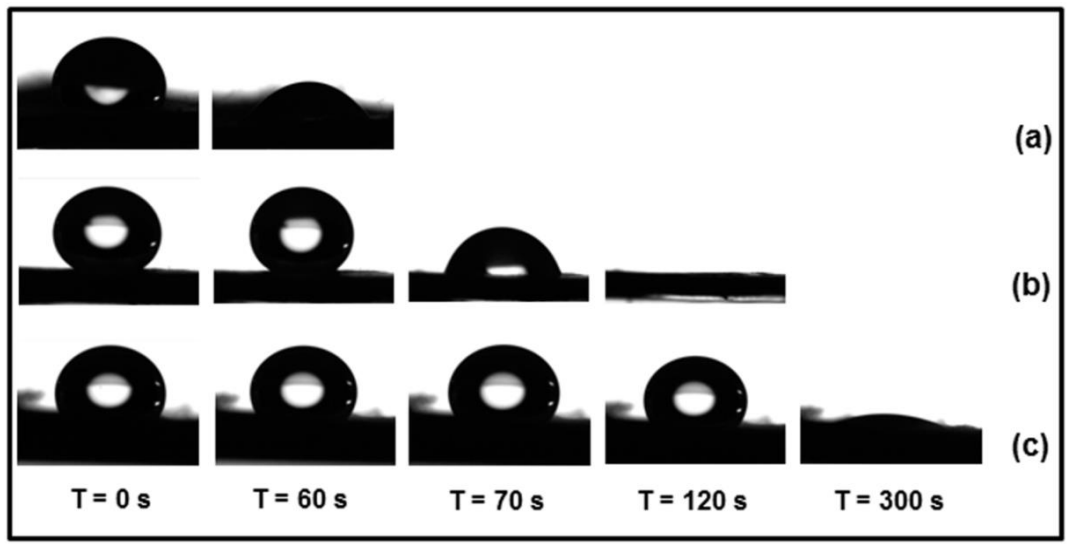

Fig. SD 6 Water adsorption process for PMAA:polyCD fibers PROP-loaded monitored by contact angle: (a) uniaxial PMAA; (b) uniaxial PMAA:polyCD (80:20); (c) uniaxial PMAA:polyCD (60:40).